\documentclass[nofootinbib,aps,preprint,10pt]{revtex4}
\usepackage{hyperref}
\usepackage{graphicx}
\usepackage{epsfig}
\usepackage{amsmath, amsfonts,euscript,bbm}
\usepackage{ifthen}
\usepackage{graphicx}
\usepackage{feyn}

\newcommand{\PRE}[1]{{#1}} % Use if preprint style

\newcommand{\roughly}[1]{\mathrel{\raise.3ex\hbox{$#1$\kern-0.85em
\lower1ex\hbox{$\sim$}}}}

\def\be{\begin{equation}}
\def\beq\begin{equation}
\def\ee{\end{equation}}
\def\bea{\begin{eqnarray}}
\def\eea{\end{eqnarray}}

\def\eqref#1{(\ref{#1})}

\def\UV{{\scriptscriptstyle U \kern-.1emV}}
\def\IR{{\scriptscriptstyle I\kern-.18em R}}

\def\gsim{\ \rlap{\raise 3pt \hbox{$>$}}{\lower 3pt \hbox{$\sim$}}\ }
\def\lsim{\ \rlap{\raise 3pt \hbox{$<$}}{\lower 3pt \hbox{$\sim$}}\ }

\begin{document}

\preprint{UCI-TR-2015-01}

\title{\PRE{\vspace*{1.3in}}
Correlation Functions of Massless Interacting Scalar Fields in de Sitter Space
\PRE{\vspace*{0.3in}}
}

\author{Arvind Rajaraman%
\PRE{\vspace*{.4in}}
}
\affiliation{Department of Physics and Astronomy, University of California, Irvine, CA  92697 USA
\PRE{\vspace*{.5in}}
}

%\date{June 2011}

\begin{abstract}
  \PRE{\vspace*{.3in}}
We examine the behavior of correlation functions
for a massless scalar field in de Sitter space with a quartic interaction. We find that
two-loop corrections are relevant, and the resummation of these corrections generates a complicated structure whereby high momentum modes
stay massless while low momentum modes develop a dynamical mass.  
\end{abstract}

\pacs{}
\maketitle

\section{Introduction}

The goal of this paper is to address the following question: what do the correlation functions
look like  for a massless field in de Sitter space with a $\phi^4$ interaction?

This question is of more than merely formal interest; it has direct connections to issues in inflationary
theory. The inflationary geometry has many similarities to de Sitter space (especially in the earlier stages of inflation). Furthermore, scalar fields play a fundamental role in inflation~\cite{Linde:2005ht},
and correlation
functions of these fields are recorded in the cosmic microwave background.

Current observations of the CMB from the WMAP experiment~\cite{Bennett:2012fp}
are currently almost exactly consistent with a scale invariant
power spectrum. However, it would be surprising if the inflaton had no
interactions whatsoever, and such interactions are expected to
produce deviations from exact scale invariance~\cite{Maldacena:2002vr}. Several experiments have been searching for such nongaussianities.
For comparison of these observations to theory, it is important
to be able to calculate correlation functions of these scalar fields
in the inflationary background.

Unfortunately, the scalar fields in inflation are typically light ($m\ll H$),
and for these small masses, it turns out that the loop corrections are important.
In fact, loop corrections involving massless fields are infrared divergent
\cite{Ford:1977in,Weinberg:2005vy,Weinberg:2006ac}.
To compare theoretical predictions to observations, it is
crucial to be able to treat these divergences rigorously. This has led to
a lot of work using many different approaches to try and
resum these divergences formally both for scalars and gravitons 
\cite{Kirsten:1993ug,Boyanovsky:2005sh,Sloth:2006az,Sloth:2006nu,Seery:2007we,
Seery:2007wf,Dimastrogiovanni:2008af,Senatore:2009cf,Tsamis:2008it,Enqvist:2008kt,Tsamis:1993ub,Kumar:2009ge,Rajaraman:2010zx,Higuchi:2009ew,vanderMeulen:2007ah,Marolf:2010zp,Marolf:2010nz,Burgess:2010dd,Starobinsky:1994bd,Riotto:2008mv,Higuchi:2010xt,{Rajaraman:2010xd}, Beneke:2012kn,Petri:2008ig, Burgess:2009bs,Boyanovsky:1998aa,Boyanovsky:2011xn,Tsamis:1996qq,Tsamis:1996qm,Hollands:2011we,Herranen:2013raa,Garbrecht:2013coa,Serreau:2013koa,Serreau:2013eoa,Gautier:2013aoa,Serreau:2013psa,Jatkar:2011ju,Garbrecht:2011gu}.
In some approaches the fact that one loop corrections are divergent for a 
massless field results in a self-consistent mass being developed. In other approaches,
the propagator becomes time dependent, breaking de Sitter invariance, and in yet other
approaches particle production leads again to an induced mass for the field.
All these techniques qualitatively lead to the result that
the  massless fields pick up a nonperturbative mass of order $\sqrt{\lambda}H$ once
interactions are included.  

 However, one would expect that a short wavelength mode of the scalar field
 should be completely unaffected by the de Sitter curvature (in much the
 same way as current lab experiments do not take into account the curvature of the
 universe). A long wavelength mode, on the other hand, can plausibly be affected by the
 curvature and may pick up a dynamical mass (indeed, this {\it must} happen in order to prevent divergences from occurring.). In short, we expect any corrections to be momentum
 dependent, with large effects only for long wavelength modes.
However, the loop effects found in the various approaches above often seem to
affect all wavelengths equally, and usually the short wavelength modes acquire a mass.

The reason for this is that the analyses in most of the above papers typically
 focus on {\it one-loop}
corrections. However, in $\phi^4$ theory, the  one-loop corrections have a very non-generic feature; they are
 {\it independent} of the external momenta. This leads to the surprising effect detailed in
the previous paragraph. This feature suggests that the one-loop corrections may not encode the relevant physics;
 indeed, all one-loop corrections in $\phi^4$ theory
can be  completely canceled by the addition of a mass counterterm.  

This further suggests that the resolution of the
divergences involves a study of the two-loop corrections in this theory.
Unlike the one-loop corrections, the 2-loop
 corrections are not independent of the external momenta.
 The long wavelength modes will then receive mass
 corrections which are
 different from the corrections to the short distance modes;
 this then allows us to cancel off the mass for the short
 distance modes using the counterterm, while maintaining a nonzero mode for the
 long distance modes which can cancel the IR divergence.

Our focus in this paper is to analyze in detail how these 2-loop corrections affect the correlation functions.
Unsurprisingly, they are much more computationally difficult than the one-loop corrections, and so
we will be unable to find closed form analytic results. However, many qualitative features can be extracted.
In particular, we show that  one loop corrections are indeed canceled by counterterms, while two-loop corrections produce a momentum dependent mass which is zero for the
short distance modes. Hence the long wavelength modes acquire a dynamical mass, while the 
short wavelength modes remain massless, as expected.

To see this effect, which is nonperturbative, we will need to resum
the two loop corrections.
Technically, we resum a subset of the two-loop diagrams which are
expected to produce the leading infrared divergences. This resummation
will be found to produce a momentum dependent effective mass.
The scaling of the masses turns out to be very different from previous calculations
in the literature.

As we shall discuss below, the difference between our results and the results in the literature
 appears to hinge on a different definition of the term 'massless scalar' (most importantly, reference \cite{Gautier:2013aoa} employs techniques extremely similar to this paper, but with a different definition of the massless limit). 
To define what we mean by this term, we 
use the general feature that the de Sitter propagator, on scales smaller
than the Hubble scale,  should locally resemble a Minkowski space propagator. 
In our 
definition, a massless field is one which for shorter length scales (i.e. distances
parametrically smaller than the Hubble scale) has a propagator similar to a 
 massless Minkowski space propagator.
This is not the only possible choice (reflecting the difference with other definitions
in the literature), but is a well defined one. The more naive definition would have been that the
propagator should approach the propagator for a massless field at low momenta, but 
this appears to run into divergences.

\section{Review of the in-in formalism}

We begin by reviewing the in-in formalism for $\phi^4$ theory. This section deals
with the free field.
These rules have already been derived and presented elsewhere
(e.g. \cite{Weinberg:2005vy, vanderMeulen:2007ah, Petri:2008ig});
we refer the interested reader to these papers for further details.

We will take the metric of de Sitter space to be
\bea
ds^2={1\over H^2\tau^2}\left(d\tau^2-\sum_{i=1}^{3}dx_i^2\right)
\eea

We will also consider a scalar field propagating in this geometry;
the field will have mass $m$ and a quartic interaction. The Lagrangian is then
\bea
{\cal L}(\phi)=\sqrt{g}\left[{1\over 2}g^{\mu\nu}\partial_\mu\phi\partial_\nu\phi
-{1\over 2}m^2\phi^2-{\lambda\over 4!}\phi^4\right]
\eea

We note that we will be focusing on very light fields.
To leading order,  therefore, we will set $m^2=0$.
Furthermore, we will have to work in the in-in formalism.
In this formalism, the number of fields is doubled; for the scalar field, we have
the fields $\phi_+,\phi_-$. We additionally define
\bea
\phi_C={1\over 2}(\phi_++\phi_-)
\qquad
\phi_\Delta=\phi_+-\phi_-
\eea
The in-in Lagrangian is then defined as
$
{\cal L}={\cal L}(\phi_+)-{\cal L}(\phi_-).
$

There are  four propagators corresponding to
the two sets of scalar fields. The Keldysh propagator is denoted $F$ and
is defined as
\bea
iF(x, y)=\langle \phi_C(x) \phi_C(y)\rangle
\eea

In addition, we have the advanced and retarded propagators
\bea
G^R(x, y)=\langle \phi_C(x) \phi_\Delta(y)\rangle&=&i\theta(x^0-y^0)(\langle \phi(x)
\phi(y)\rangle-\langle \phi(y) \phi(x)\rangle)
\\
G^A(x, y)&=&G^R(y, x)
\eea

The fourth propagator $\langle \phi_\Delta(x) \phi_\Delta(y)\rangle$ is identically
zero.

In Feynman diagrams, $\phi_C$ is denoted by a solid line, and $\phi_\Delta$ by
a dashed line. The propagators (after a Fourier transform) are denoted by
\bea
\feyn{\vertexlabel^{\tau_1} {f f }\vertexlabel^{\tau_2}}~~~~~~=~~~~~~~F(k,\tau_1,\tau_2)
\\
\feyn{\vertexlabel^{\tau_1} {f h }\vertexlabel^{\tau_2}}~~~~~~=~-iG^R(k,\tau_1,\tau_2)
\eea

There are two vertices in $\phi^4$ theory.

\bea
\Diagram{hd hu
\\
hu fd}=-i{\lambda\over 4}a^4(\tau)
\\
\Diagram{fd fu
%\\
%\hskip -0.2 cm f h
\\
fu hd}=-i\lambda a^4(\tau)
\eea

We can calculate these propagators in the free field limit i.e. when $\lambda=0$.
For free fields, the propagators can be found to be (we will denote the
free field limit by a 0 superscript)\cite{vanderMeulen:2007ah}
\bea
F^{(0)}_{m^2}(k,\tau_1,\tau_2)={\pi H^2\over 4}(\tau_1\tau_2)^{3/2}Re(H_\nu(-k\tau_1)H^*_\nu(-k\tau_2))
\\
G^{R(0)}_{m^2}(k,\tau_1,\tau_2)=-\theta(\tau_1-\tau_2){\pi H^2\over 2}
(\tau_1\tau_2)^{3/2}Im(H_\nu(-k\tau_1)H^*_\nu(-k\tau_2))
\eea
where $\nu^2={9\over 4}-{m^2\over H^2}$.

There are several limits of the free field $F$-propagator that are of interest. The first
is the zero mass limit. 
For zero mass, we have
 \bea
F^{(0)}_{m^2=0}(k,\tau_1,\tau_2)= {H^2\over 2k^3}[(1+k^2\tau_1\tau_2)\cos(k(\tau_1-\tau_2))
+k(\tau_1-\tau_2)\sin(k(\tau_1-\tau_2))]~~~~~~~~~~~~~~~~
\\
G^{R(0)}_{m^2=0}(k,\tau_1,\tau_2)= \theta(\tau_1-\tau_2){H^2\over k^3}
[(1+k^2\tau_1\tau_2)\sin(k(\tau_1-\tau_2))
-k(\tau_1-\tau_2)\cos(k(\tau_1-\tau_2))]
\eea

 For small $k$ the  Keldysh propagator goes as ${1\over k^3}$
 \bea
F^{(0)}_{m^2=0}(k,\tau_1,\tau_2)\sim {H^2\over 2k^3}~~~~~~~~~~~~~~~~
\eea
This leads to divergences in loop diagrams.
For finite masses, the low momentum behavior of the propagator is
expressible in terms of
${m^2\over 3H^2}=\epsilon\ll 1$ and is found to be
\bea
F^{(0)}_{m^2}(k,\tau_1,\tau_2)\sim {H^2\over 2k^{3}}(k^2\tau_1\tau_2)^{\epsilon}
\eea
which regulates the infrared divergences found at $m^2=0$.

\section{ One-loop corrections}

We first briefly consider the one-loop corrections to the massless theory.
The relevant diagrams are shown below. The third and fourth diagrams represent counterterms.
\begin{figure} [h]
\begin{center}
\includegraphics[width=1.0\textwidth,angle=0]{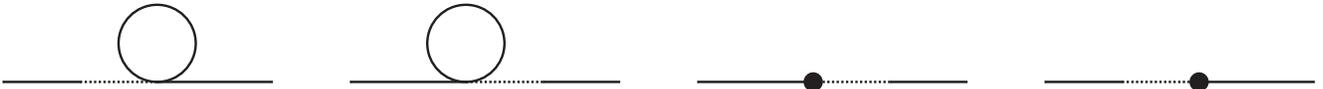}
\caption{One loop corrections}
\end{center}
\label{Oneloop}
\end{figure}

In a perturbation expansion, the first diagram is of the form
\bea
-i{\lambda\over 4}a^4(\tau_3)G^{(0)}(k,\tau_1,\tau_3)\left[\int d^3q
F^{(0)}(q,\tau_3,\tau_3)\right]F^{(0)}(k,\tau_3,\tau_2)
\eea

For $m^2=0$, $F^{(0)}(q,\tau_3,\tau_3)\sim {1\over q^3}$ at small $q$, and so the
integral is divergent. The massless theory therefore does not
seem to have  a good perturbation expansion. 
However, the infinite correction can be canceled by a suitable choice of the counterterms.
Since  the loop diagram is independent of external momenta, 
canceling the mass at high energies also cancels the mass 
at zero momentum.
 This will again result
in a massless zero mode which will lead to IR divergences.
We must therefore go beyond one loop to see whether the divergences 
can be resolved.

\section{ Two-loop corrections}
 
 The
two loop correction has the generic form of a sunrise diagram.
There are in fact three separate two-loop corrections which occur in this
theory, which we denote $\Sigma_{1,2,3}$.

 The first set of corrections is shown in Figure 2.
\begin{figure} [h]
\begin{center}
\includegraphics[width=0.8\textwidth,angle=0]{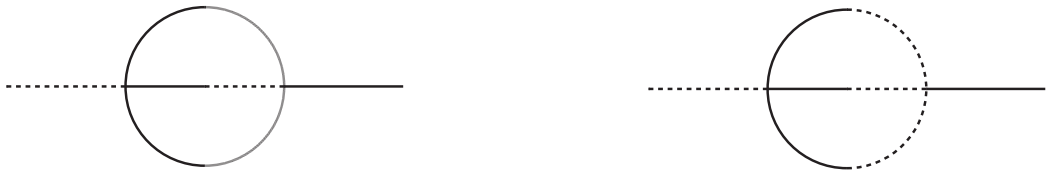}
\caption{Corrections contributing to $\Sigma_1$}
\end{center}
\end{figure}
These corrections can be written explicitly as
\bea
\Sigma_1(k,\tau_1,\tau_2)=i{\lambda^2}a^4(\tau_1)a^4(\tau_2)\int d^3k_1 \int d^3k_2
F(k_1,\tau_1,\tau_2)F(k_2,\tau_1,\tau_2)
G^R(k-k_1-k_2,\tau_1,\tau_2)\nonumber\\
-i{\lambda^2\over 4}a^4(\tau_1)a^4(\tau_2)\int d^3k_1 \int d^3k_2 G^R(k_1,\tau_1,\tau_2)
G^R(k_2,\tau_1,\tau_2)
G^R(k-k_1-k_2,\tau_1,\tau_2)
\label{Sigmaintegral}
\eea

This loop can be thought of as inducing a new
vertex corresponding to a (nonlocal) term in the Lagrangian
proportional to $\phi_C\phi_\delta$.

There is an analogous vertex where
we reverse the time direction; this diagram is the mirror
image of the diagram above. This mirrored vertex will
be denoted $\Sigma_2$ and can be thought of as inducing a new
vertex corresponding to a (nonlocal) term in the Lagrangian
proportional to $\phi_\delta\phi_C$.

Finally, there is a third set of loop
corrections which induce a new
vertex corresponding to a (nonlocal) term in the Lagrangian
proportional to $\phi_\Delta\phi_\Delta$.
This set of diagrams will be denoted $\Sigma_3$, and can be written as
\bea
\Sigma_3(k,\tau_1,\tau_2)={\lambda^2\over 4}a^4(\tau_1)a^4(\tau_2)
\int d^3k_1 \int d^3k_2 G^R(k_1,\tau_1,\tau_2)G^R(k_2,\tau_1,\tau_2)
F(k-k_1-k_2,\tau_1,\tau_2)\\
+{\lambda^2\over 4}a^4(\tau_1)a^4(\tau_2)\int d^3k_1 \int d^3k_2
G^A(k_1,\tau_1,\tau_2)G^A(k_2,\tau_1,\tau_2)
F(k-k_1-k_2,\tau_1,\tau_2)\\
-{\lambda^2}a^4(\tau_1)a^4(\tau_2)\int d^3k_1 \int d^3k_2 F(k_1,\tau_1,\tau_2)F(k_2,\tau_1,\tau_2)
F(k-k_1-k_2,\tau_1,\tau_2)
\eea

For compactness, we shall replace all loops $\Sigma_i$ in Feynman
diagrams by a
insertion labeled with the appropriate index i.
Therefore for example, we shall indicate a $\Sigma_3$ insertion by
a dot with a 3 subscript as shown in figure 3.

\begin{figure} [h]
\begin{center}
\includegraphics[width=0.6\textwidth,angle=0]{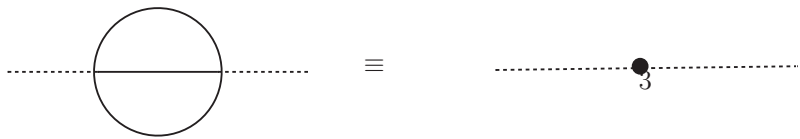}
\caption{A condensed notation for the loop integrals}
\end{center}
\end{figure}

\section{The high momentum region}

These diagrams are difficult to compute exactly. We are, however, interested
in the most divergent pieces of these diagrams; this will allow us to make several
approximations to obtain an estimate of these loop corrections.
We will focus on $\Sigma_1$ at the beginning, and apply the same approximations
to the other integrals afterward.

We first note that the IR divergences primarily
come from regions of the loop integral where one
of the $F$ propagators has a small momentum flowing through it.
Indeed, if we were to
replace $F$ by $F^{(0)}$ in the above integrals, they would all be infrared divergent because
$F^{(0)}(k,\tau_1,\tau_2)\sim {1\over k^3}$.
This IR divergence is presumably regulated when the full $F$ propagator is used.
We therefore conclude that the low momentum behavior of the full $F$ propagator is significantly different
from the tree level propagator $F^0$. 
This modification by the higher order
 corrections will moderate the infrared
 behavior and make the integrals finite.
 Now, although the $F$ propagator is modified in such a way that
 the integrals are finite, the divergences must reappear if we
 take the interaction $\lambda\to 0$.
 It must therefore be the case that we get a large contribution (i.e.
 parametrically larger as a function of $\lambda$) from the region of integration where the
 momenta flowing through an $F$ propagator is small. This further indicates that we can keep 
only the terms with the largest
number of $F$ propagators, which are expected to have the largest contributions.

On the other hand, the $G$
propagator does not have infrared divergences when
the momentum flowing through it is small. This means that there
is no requirement for the corrections to
 modify it significantly at low momenta. We will assume the corrections
 to $G$ are small (i.e. suppressed by powers of $\lambda$) which will be
 confirmed by later calculations. We will therefore (to leading
order) replace $G$ by $G^0$ in the above expressions.

Furthermore, the integration over momenta is expected to be dominated by the regions when 
both $F$ propagators have
small momenta in their arguments. In particular, we would expect the dominant part of
the integral for $\Sigma_1$ to come from the integration region where both internal momenta $k_1, k_2$
are much smaller than the external momentum $k$ i.e.
$k_1,k_2 \ll k$. 
The integral then  simplifies to
\bea
\Sigma_1(k,\tau_1,\tau_2)\simeq i%{\lambda^2}
a^4(\tau_1)a^4(\tau_2)G^{R(0)}(k,\tau_1,\tau_2)\times [c(k,\tau_1,\tau_2)]^2
\eea
where we have defined
\bea
c(k,\tau_1,\tau_2)=
\lambda\int_0^k d^3k_1 F(k_1,\tau_1,\tau_2)\label{highmomentummass}
\eea

The next point to note is that the leading term in
 the $c(k,\tau_1,\tau_2)$  integral above
is expected to be time independent.  The integral diverges if $F$ is replaced
by  the tree level $F^0$ propagator, which for small $k$ leads to an integral of
the form ${d^3k\over k^3}$. The divergent term is time independent.
When the divergence is regulated,
the integral will be large but finite;
it is reasonable to expect
that the leading value (in an expansion in $\lambda$) will continue to be time independent.
We will defined this leading term to be $m^2_1(k)$.

With these approximations, we are able to simplify the $\Sigma_1$ integral to
\bea
\Sigma_1(k,\tau_1,\tau_2)\simeq ia^4(\tau_1)a^4(\tau_2)m_1^4(k)
G^{R(0)}(k,\tau_1,\tau_2)
\eea

We can now resum the two loop corrections.
For the retarded propagator, the corrections that we
wish to sum are  of the form shown in Figure 4.
\begin{figure} [h]
\begin{center}
\includegraphics[width=0.45\textwidth,angle=0]{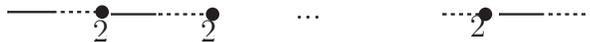}
\caption{Corrections to the retarded propagator}
\end{center}
\end{figure}

 These lead to an equation for the full
propagator
 \bea
G^{R}(k,\tau_1,\tau_2) =G^{R(0)}(k,\tau_1,\tau_2)~~~~~~~~~~~~~~~~~~~~~~~~~~~~~~~~~~~~~~~~~~~
 ~~~~~~~~~~~~~~~~~~~~~~~~~~~~~~~~
 \\
 +
 m_1^4(k)\int {d\tau_3\over (H\tau_3)^4}{d\tau_4\over (H\tau_4)^4}
 G^{R(0)}(k,\tau_1,\tau_3)G^{R(0)}(k,\tau_3,\tau_4)G^{R(0)}(k,\tau_4,\tau_2)~~~~~~~~~~~~~~~~~
 ~~~
 \\
 + m_1^8(k)\int {d\tau_3\over (H\tau_3)^4}{d\tau_4\over (H\tau_4)^4}
{d\tau_5\over (H\tau_5)^4}{d\tau_6\over (H\tau_6)^4} G^{(0)}(k,\tau_1,\tau_3)
G^{R(0)}(k,\tau_3,\tau_4)G^{R(0)}(k,\tau_4,\tau_5)
\\
\times G^{R(0)}(k,\tau_5,\tau_6)
 G^{R(0)}(k,\tau_6,\tau_2)
 \\
 +~~~~~~...~~~~~~~~~~~~~~~~~~~~~~~~~~~~~~~~~~~~~~~~~~~~~~~~~~~~~~~~~~~~~~~~~~
 ~~~~~~~~~~~~~~~~~~~
 \eea

We can perform the sum by comparing it to the expansion
when a mass term is treated as a perturbation.
A massive field has the exact retarded propagator 
$ G_{m^2}(k,\tau_1,\tau_2)$.
On the other hand, in the same theory, the mass term $-m^2\phi_C\phi_\Delta$ can be treated as a perturbation
to the massless theory. 
We can match the
exact solution to the perturbation expansion; this
yields
 \bea
 \label{masspert}
 G^{R(0)}_{m^2}(k,\tau_1,\tau_2) =G_{m^2=0}^{R(0)}(k,\tau_1,\tau_2)~~~~~~~~~~~~~
 ~~~~~~~~~~~~~~~~~~~~~~~~~~~~~~~~~~~~~~~~~~~~~~~~~~
 \\
 -\ m^2 \int {d\tau_3\over (H\tau_3)^4} G_{m^2=0}^{R(0)}(k,\tau_1,\tau_3)G_{m^2=0}^{R(0)}(k,\tau_3,\tau_2)
~~~~~~~~~~~~~~~~~~~~~~~~~~~
 \\
 +\ m^4\int {d\tau_3\over (H\tau_3)^4}{d\tau_4\over (H\tau_4)^4}
 G_{m^2=0}^{R(0)}(k,\tau_1,\tau_3)G_{m^2=0}^{R(0)}(k,\tau_3,\tau_4)G_{m^2=0}^{R(0)}(k,\tau_4,\tau_2)
 \\
 +~~~~~~...~~~~~~~~~~~~~~~~~~~~~~~~~~~~~~~~~~~~~~~~~~~~~~~~~~~~~~~~~~~~~~~~~~~~~~~~
 \eea

 Comparing this to the previous expansion involving $\Sigma_1$, we find
\bea
 G^{R}(k,\tau_1,\tau_2)={1\over 2}
(G^{R(0)}_{m_1^2(k)}(k,\tau_1,\tau_2)+G^{R(0)}_{-m_1^2(k)}(k,\tau_1,\tau_2))
\eea

\begin{figure} [h]
\begin{center}
\includegraphics[width=0.8\textwidth,angle=0]{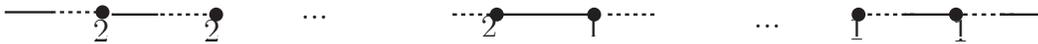}
\caption{Corrections to the Keldysh propagator}
\end{center}
\end{figure}

 Similarly, the $F$ propagator receives
 contributions of the form shown in Figure 5.
Just like the  the retarded propagator, we find that the contribution of these terms
is ${1\over 2}(F_{m_1^2(k)}(k,\tau_1,\tau_2)+F_{-m_1^2(k)}(k,\tau_1,\tau_2))$.

We also need to consider contributions which involve a $\Sigma_3$ vertex insertion. These are all
of the form shown in Fig. 6.
\begin{figure} [h]
\begin{center}
\includegraphics[width=0.8\textwidth,angle=0]{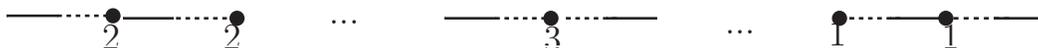}
\caption{Other corrections to the Keldysh propagator}
\end{center}
\end{figure}

Comparing to the expansion for the retarded propagator, we find that these contributions
can be written as
\bea\int {d\tau_3\over (H\tau_3)^4}{d\tau_4\over (H\tau_4)^4}G^R (k,\tau_1,\tau_3)
\Sigma_3(k,\tau_3,\tau_4) G^A(k,\tau_4,\tau_2)\eea
 We shall treat these diagrams in perturbation theory; that is, we
shall not include them in the resummation.
We will hence ignore these terms for the remainder of the discussion.
It would be interesting to examine the effects of these diagrams; we shall leave
this for future work.

Summarizing our results so far,  we have found that the resummed Keldysh propagator
is of the form 
\bea
F(k,\tau_1,\tau_2)={1\over 2}(F_{m_1^2(k)}(k,\tau_1,\tau_2)+F_{-m_1^2(k)}(k,\tau_1,\tau_2))
\label{highmomentumF}\eea
where  $m^2_1(k)$ is the leading term in the integral (\ref{highmomentummass}).
Note that this is {\it not} the propagator for a massive field. 

\section{The Low Momentum Region}

Unfortunately, our result above cannot hold for all external momenta.
For if it did, (\ref{highmomentumF}) would be the exact solution for the propagator.
But now the term in the propagator  $F_{-m_1^2(k)}(k,\tau_1,\tau_2)$ grows in the infrared even faster than
$F_{0}(k,\tau_1,\tau_2)$. From
(\ref{highmomentummass}), we would find
 $m^2_1(k)$ to be divergent. This implies that the solution is not self consistent.

One possibility that there is simply no solution; that there is a deep inconsistency
that does not allow a propagator to exist when the one-loop corrections are canceled by the counterterms.
While this possibility cannot be neglected, it would be surprising, for the
reasons outlined in the introduction.
 We therefore conclude that some other corrections must become relevant at low momentum. 
The consequence is that there must be a crossover momentum $k_0$ below which our solution 
(\ref{highmomentumF}) breaks down.

 We can  trace this breakdown back to the original simplifications
 we made for the integral.
We found that the leading contribution to $\Sigma_1$ was
\bea
\Sigma_1(k,\tau_1,\tau_2)\simeq i{\lambda^2}
a^4(\tau_1)a^4(\tau_2)\int d^3k_1 \int d^3k_2 F(k_1,\tau_1,\tau_2)F(k_2,\tau_1,\tau_2)
G^{R(0)}(k-k_1-k_2,\tau_1,\tau_2)
\eea

We assumed that the main contribution came from the region of integration where
the momenta $k_1, k_2$ were very small, and in particular much smaller than the external
momentum $k$. Under this assumption, 
the integral was found to be proportional to $[c(k,\tau_1,\tau_2)]^2$ where $c(k,\tau_1,\tau_2)$ was  
the integral
of $F(k_1,\tau_1,\tau_2)$ up to momentum $k$. 
But for small external momenta $k$, the last factor necessarily goes to zero. It therefore cannot be dominant
for small external momenta. Once again we are led to the existence
of a crossover momentum $k_0$ below which the solution (\ref{highmomentumF}) is not valid.

On the other hand, for larger momenta ($k$ above $k_0$), our resummation in the previous section is still valid.
We then expect the form of the propagator at these momenta to be of the form
(\ref{highmomentumF}).

For  external momenta which are smaller than $k_0$, 
we can approximate the integral for $\Sigma_1$
by its value at $k=0$:
   \bea
\Sigma_1(k,\tau_1,\tau_2)\simeq i{\lambda^2}a^4(\tau_1)a^4(\tau_2)\int_{k_1,k_2=0}^{\infty} d^3k_1 d^3k_2 F(k_1,\tau_1,\tau_2)F(k_2,\tau_1,\tau_2)G^{R}(-k_1-k_2,\tau_1,\tau_2)
\label{lowmomentummass}
\eea
This is independent of $k$, and
hence produces a correction that is dominant for low momenta. These corrections
will therefore produce an effect similar to a dynamical mass, and in particular,
they will cut off the infrared divergences.
The time dependence is more complicated than a simple mass insertion, 
and so we therefore make the ansatz
\bea
F(k,\tau_1,\tau_2)=
\left\{
\begin{array}{c}
{1\over 2}(F_{m_1^2(k)}(k,\tau_1,\tau_2)+F_{-m_1^2(k)}(k,\tau_1,\tau_2))
\qquad k>k_0
\\
f(\tau_1,\tau_2){H^2\over 2k^{3}}(k^2)^{\epsilon'}
\qquad k<k_0
\end{array}
\right.
\label{finalresult}
\eea
where  ${m_1^2\over 3H^2}=\epsilon, {m_2^2\over 3H^2}=\epsilon'$, and we have parametrized
the time dependence at low momenta by the function $f$.

We now can attempt to find the scaling of the various parameters of
our ansatz with $\lambda$.
Comparing the equations (\ref{highmomentummass}), which contains one integral over $\lambda F$,
 and
(\ref{lowmomentummass}), which contains two integrals over $\lambda F$, we expect 
 $m_2^2\sim (m^2_1)^2$, and thus that $m_2^2$ is parametrically smaller 
than $m_1^2$. In turn, this means that the
integrals in these equations are dominated by the low
momenta regions $k<k_0$.
We then find
\bea
m^2_1(k) \propto  \lambda (k_0^2)^{-\epsilon}({1\over {2\epsilon'}})
\qquad \qquad
m_2^2\sim \lambda^2(k_0^2)^{-2\epsilon}({1\over {\epsilon'}^2})
\eea
Since $\epsilon'$ is parametrically small in $\lambda$, we can set
to leading order $(k_0^2)^{-2\epsilon}=1$.
We then find $m_2^2\sim \lambda^{2/3}$ and $m_1^2\sim \lambda^{1/3}$.

Our conclusion for the low momentum modes is then that their propagation is
modified in a manner similar to a dynamical mass. This dynamical mass has a scaling proportional
to $\lambda^{1/3}$ and cuts off the infrared divergences.

\section{Discussion and Conclusion}

We have discussed how a quartic interaction would modify the propagator of a massless field in de Sitter space.
We have found a complicated result; there is a crossover momentum scale with very different 
behavior of the propagator above and below the crossover. For long wavelengths,
the modes behave as if they develop a dynamical mass. For shorter  wavelengths,
surprisingly, the propagator is the sum  of  a massive propagator and
a propagator for a tachyonic field. 
Our main result is encapsulated in equation (\ref{finalresult}), which is our
result for the Keldysh propagator once interactions are included.

The apparently surprising behavior at shorter momenta
 has a natural explanation from the masslessness of the scalar.
Consider the behavior of the free Keldysh propagator for
$\tau_1=\tau_2=\tau$ with $k\tau$ large.
The free propagator has the  expansion
\bea
F^{(0)}_{m^2}(k,\tau_1,\tau_2)= {H^2\tau^2\over 2k}[1+{1\over k^2\tau^2}+{m^2\over 2k^2H^2\tau^2} )+...]
\eea
We can therefore define the mass of the particle by looking at the deformation
away from the massless propagator i.e.
\bea
m^2= \lim_{k\rightarrow \infty} 4k^3[F(k,\tau_1,\tau_2)-F^{(0)}_{m^2=0}(k,\tau_1,\tau_2)]
\eea

For the expression (\ref{highmomentumF}), we find that the mass is zero, thereby justifying calling
this a massless scalar.
At this order in the loop expansion, the high momentum modes in our calculation
do not obtain a mass
correction. Note that the tachyonic piece was necessary for this to happen.

The behavior for longer wavelengths is more in line with expectations.
We have found that momenta with longer wavelengths behave as if they have a mass,
which is similar to results in the literature. 
Quantitatively, however, our results are different;
 we have found a dynamical mass which 
scales as $\lambda^{1/3}$ rather than the usual scaling of
$\lambda^{1/4}$ found by previous authors. This is because the
mass in our case is generated at two loops rather than one-loop.
 
Further corrections to the propagator and other correlation functions can
be calculated in perturbation theory. Infrared divergences are now
absent and so the perturbation expansion should make sense. We note that any loop
involving an $F$ propagator is enhanced by a factor ${1\over m_2^2}\sim \lambda^{-2/3}$.
This indicates that the perturbation expansion is in powers of $\lambda^{1/3}$ rather than $\lambda$.

There are several open issues that still need to be addressed. In particular, we have argued for a crossover momentum, but we have not found its magnitude (even as a scaling in $\lambda$). 
A calculation to next order in perturbation theory might shed light on this issue. We shall leave this 
and other questions for future work. 

\section{Acknowledgments}

This work was supported in part by  NSF grant PHY-1316792.


\begin{thebibliography}{99}

%\cite{Linde:2005ht}
\bibitem{Linde:2005ht} 
  A.~D.~Linde,
  %``Particle physics and inflationary cosmology,''
  Contemp.\ Concepts Phys.\  {\bf 5}, 1 (1990)
  [hep-th/0503203].
  %%CITATION = HEP-TH/0503203;%%

%\cite{Bennett:2012fp}
\bibitem{Bennett:2012fp} 
  C.~L.~Bennett, D.~Larson, J.~L.~Weiland, N.~Jarosik, G.~Hinshaw, N.~Odegard, K.~M.~Smith and R.~S.~Hill {\it et al.},
  %``Nine-Year Wilkinson Microwave Anisotropy Probe (WMAP) Observations: Final Maps and Results,''
  arXiv:1212.5225 [astro-ph.CO].
  %%CITATION = ARXIV:1212.5225;%%
	
	\bibitem{Maldacena:2002vr}
Juan~Martin Maldacena.
\newblock Non-gaussian features of primordial fluctuations in single field
  inflationary models.
\newblock {\em JHEP}, 05:013, 2003.
	
	
	%\cite{Weinberg:2005vy}
\bibitem{Weinberg:2005vy} 
  S.~Weinberg,
  %``Quantum contributions to cosmological correlations,''
  Phys.\ Rev.\ D {\bf 72}, 043514 (2005)
  [hep-th/0506236].
  %%CITATION = HEP-TH/0506236;%%	
	
	%\cite{Weinberg:2006ac}
\bibitem{Weinberg:2006ac} 
  S.~Weinberg,
  %``Quantum contributions to cosmological correlations. II. Can these corrections become large?,''
  Phys.\ Rev.\ D {\bf 74}, 023508 (2006)
  [hep-th/0605244].
  %%CITATION = HEP-TH/0605244;%%
	
		%\cite{Ford:1977in}
\bibitem{Ford:1977in} 
  L.~H.~Ford and L.~Parker,
  %``Infrared Divergences in a Class of Robertson-Walker Universes,''
  Phys.\ Rev.\ D {\bf 16}, 245 (1977).
  %%CITATION = PHRVA,D16,245;%%
	
	
	
	%\cite{Kirsten:1993ug}
\bibitem{Kirsten:1993ug} 
  K.~Kirsten and J.~Garriga,
  %``Massless minimally coupled fields in de Sitter space: O(4) symmetric states versus de Sitter invariant vacuum,''
  Phys.\ Rev.\ D {\bf 48}, 567 (1993)
  [gr-qc/9305013].
  %%CITATION = GR-QC/9305013;%%
	

	
	%\cite{Boyanovsky:2005sh}
\bibitem{Boyanovsky:2005sh} 
  D.~Boyanovsky, H.~J.~de Vega and N.~G.~Sanchez,
  %``Quantum corrections to slow roll inflation and new scaling of superhorizon fluctuations,''
  Nucl.\ Phys.\ B {\bf 747}, 25 (2006)
  [astro-ph/0503669].
  %%CITATION = ASTRO-PH/0503669;%%
	
	%\cite{Sloth:2006az}
\bibitem{Sloth:2006az} 
  M.~S.~Sloth,
  %``On the one loop corrections to inflation and the CMB anisotropies,''
  Nucl.\ Phys.\ B {\bf 748}, 149 (2006)
  [astro-ph/0604488].
  %%CITATION = ASTRO-PH/0604488;%%
	
	%\cite{Sloth:2006nu}
\bibitem{Sloth:2006nu} 
  M.~S.~Sloth,
  %``On the one loop corrections to inflation. II. The Consistency relation,''
  Nucl.\ Phys.\ B {\bf 775}, 78 (2007)
  [hep-th/0612138].
  %%CITATION = HEP-TH/0612138;%%
	
	%\cite{Seery:2007we}
\bibitem{Seery:2007we} 
  D.~Seery,
  %``One-loop corrections to a scalar field during inflation,''
  JCAP {\bf 0711}, 025 (2007)
  [arXiv:0707.3377 [astro-ph]].
  %%CITATION = ARXIV:0707.3377;%%
	
	%\cite{Seery:2007wf}
\bibitem{Seery:2007wf} 
  D.~Seery,
  %``One-loop corrections to the curvature perturbation from inflation,''
  JCAP {\bf 0802}, 006 (2008)
  [arXiv:0707.3378 [astro-ph]].
  %%CITATION = ARXIV:0707.3378;%%
	
	%\cite{Dimastrogiovanni:2008af}
\bibitem{Dimastrogiovanni:2008af} 
  E.~Dimastrogiovanni and N.~Bartolo,
  %``One-loop graviton corrections to the curvature perturbation from inflation,''
  JCAP {\bf 0811}, 016 (2008)
  [arXiv:0807.2790 [astro-ph]].
  %%CITATION = ARXIV:0807.2790;%%
	
	%\cite{Senatore:2009cf}
\bibitem{Senatore:2009cf} 
  L.~Senatore and M.~Zaldarriaga,
  %``On Loops in Inflation,''
  JHEP {\bf 1012}, 008 (2010)
  [arXiv:0912.2734 [hep-th]].
  %%CITATION = ARXIV:0912.2734;%%
	

	
	%\cite{Tsamis:2008it}
\bibitem{Tsamis:2008it} 
  N.~C.~Tsamis and R.~P.~Woodard,
  %``A Simplified Quantum Gravitational Model of Inflation,''
  Class.\ Quant.\ Grav.\  {\bf 26}, 105006 (2009)
  [arXiv:0807.5006 [gr-qc]].
  %%CITATION = ARXIV:0807.5006;%%
	

	
	%\cite{Enqvist:2008kt}
\bibitem{Enqvist:2008kt} 
  K.~Enqvist, S.~Nurmi, D.~Podolsky and G.~I.~Rigopoulos,
  %``On the divergences of inflationary superhorizon perturbations,''
  JCAP {\bf 0804}, 025 (2008)
  [arXiv:0802.0395 [astro-ph]].
  %%CITATION = ARXIV:0802.0395;%%
	
	%\cite{Tsamis:1993ub}
\bibitem{Tsamis:1993ub} 
  N.~C.~Tsamis and R.~P.~Woodard,
  %``The Physical basis for infrared divergences in inflationary quantum gravity,''
  Class.\ Quant.\ Grav.\  {\bf 11}, 2969 (1994).
  %%CITATION = CQGRD,11,2969;%%
	
	
	
	%\cite{Kumar:2009ge}
\bibitem{Kumar:2009ge} 
  J.~Kumar, L.~Leblond and A.~Rajaraman,
  %``Scale Dependent Local Non-Gaussianity from Loops,''
  JCAP {\bf 1004}, 024 (2010)
  [arXiv:0909.2040 [astro-ph.CO]].
  %%CITATION = ARXIV:0909.2040;%%
	
	%\cite{Rajaraman:2010zx}
\bibitem{Rajaraman:2010zx} 
  A.~Rajaraman, J.~Kumar and L.~Leblond,
  %``Constructing Infrared Finite Propagators in Inflating Space-time,''
  Phys.\ Rev.\ D {\bf 82}, 023525 (2010)
  [arXiv:1002.4214 [hep-th]].
  %%CITATION = ARXIV:1002.4214;%%
  
  %\cite{Higuchi:2009ew}
\bibitem{Higuchi:2009ew} 
  A.~Higuchi and Y.~C.~Lee,
  %``Conformally-coupled massive scalar field in de Sitter expanding universe with the mass term treated as a perturbation,''
  Class.\ Quant.\ Grav.\  {\bf 26}, 135019 (2009)
  [arXiv:0903.3881 [gr-qc]].
  %%CITATION = ARXIV:0903.3881;%%
	
		%\cite{vanderMeulen:2007ah}
\bibitem{vanderMeulen:2007ah} 
  M.~van der Meulen and J.~Smit,
  %``Classical approximation to quantum cosmological correlations,''
  JCAP {\bf 0711}, 023 (2007)
  [arXiv:0707.0842 [hep-th]].
  %%CITATION = ARXIV:0707.0842;%%
  
  	%\cite{Marolf:2010zp}
\bibitem{Marolf:2010zp} 
  D.~Marolf and I.~A.~Morrison,
  %``The IR stability of de Sitter: Loop corrections to scalar propagators,''
  Phys.\ Rev.\ D {\bf 82}, 105032 (2010)
  [arXiv:1006.0035 [gr-qc]].
  %%CITATION = ARXIV:1006.0035;%%
	
	%\cite{Marolf:2010nz}
\bibitem{Marolf:2010nz} 
  D.~Marolf and I.~A.~Morrison,
  %``The IR stability of de Sitter QFT: results at all orders,''
  Phys.\ Rev.\ D {\bf 84}, 044040 (2011)
  [arXiv:1010.5327 [gr-qc]].
  %%CITATION = ARXIV:1010.5327;%%
  
  	%\cite{Burgess:2010dd}
\bibitem{Burgess:2010dd} 
  C.~P.~Burgess, R.~Holman, L.~Leblond and S.~Shandera,
  %``Breakdown of Semiclassical Methods in de Sitter Space,''
  JCAP {\bf 1010}, 017 (2010)
  [arXiv:1005.3551 [hep-th]].
  %%CITATION = ARXIV:1005.3551;%%

	
	%\cite{Starobinsky:1994bd}
\bibitem{Starobinsky:1994bd} 
  A.~A.~Starobinsky and J.~Yokoyama,
  %``Equilibrium state of a selfinteracting scalar field in the De Sitter background,''
  Phys.\ Rev.\ D {\bf 50}, 6357 (1994)
  [astro-ph/9407016].
  %%CITATION = ASTRO-PH/9407016;%%
	
	%\cite{Riotto:2008mv}
\bibitem{Riotto:2008mv} 
  A.~Riotto and M.~S.~Sloth,
  %``On Resumming Inflationary Perturbations beyond One-loop,''
  JCAP {\bf 0804}, 030 (2008)
  [arXiv:0801.1845 [hep-ph]].
  %%CITATION = ARXIV:0801.1845;%%
  
  %\cite{Higuchi:2010xt}
\bibitem{Higuchi:2010xt} 
  A.~Higuchi, D.~Marolf and I.~A.~Morrison,
  %``On the Equivalence between Euclidean and In-In Formalisms in de Sitter QFT,''
  Phys.\ Rev.\ D {\bf 83}, 084029 (2011)
  [arXiv:1012.3415 [gr-qc]].
  %%CITATION = ARXIV:1012.3415;%%
	
		
	
	%\cite{Rajaraman:2010xd}
\bibitem{Rajaraman:2010xd} 
  A.~Rajaraman,
  %``On the proper treatment of massless fields in Euclidean de Sitter space,''
  Phys.\ Rev.\ D {\bf 82}, 123522 (2010)
  [arXiv:1008.1271 [hep-th]].
  %%CITATION = ARXIV:1008.1271;%%
	
	%\cite{Beneke:2012kn}
\bibitem{Beneke:2012kn}
  M.~Beneke and P.~Moch,
  %``On “dynamical mass” generation in Euclidean de Sitter space,''
  Phys.\ Rev.\ D {\bf 87}, no. 6, 064018 (2013)
  [arXiv:1212.3058].
  %%CITATION = ARXIV:1212.3058;%%
	

\bibitem{Petri:2008ig}
G.~Petri.
\newblock {A Diagrammatic Approach to Scalar Field Correlators during
  Inflation}.
\newblock 2008.

\bibitem{Burgess:2009bs}
C.P. Burgess, L.~Leblond, R.~Holman, and S.~Shandera.
\newblock {Super-Hubble de Sitter Fluctuations and the Dynamical RG}.
\newblock {\em JCAP}, 1003:033, 2010.



\bibitem{Boyanovsky:1998aa}
D.~Boyanovsky, H.J. de~Vega, R.~Holman, and M.~Simionato.
\newblock {Dynamical renormalization group resummation of finite temperature
  infrared divergences}.
\newblock {\em Phys.Rev.}, D60:065003, 1999.



\bibitem{Boyanovsky:2011xn}
Daniel Boyanovsky and Richard Holman.
\newblock {On the Perturbative Stability of Quantum Field Theories in de Sitter
  Space}.
\newblock {\em JHEP}, 1105:047, 2011.


\bibitem{Tsamis:1996qq}
N.~C. Tsamis and R.~P. Woodard.
\newblock {Quantum Gravity Slows Inflation}.
\newblock {\em Nucl. Phys.}, B474:235--248, 1996.

\bibitem{Tsamis:1996qm}
N.~C. Tsamis and R.~P. Woodard.
\newblock {The quantum gravitational back-reaction on inflation}.
\newblock {\em Annals Phys.}, 253:1--54, 1997.


\bibitem{Hollands:2011we}
Stefan Hollands.
\newblock {Massless interacting quantum fields in deSitter spacetime}.
\newblock 2011.


%\cite{Herranen:2013raa}
\bibitem{Herranen:2013raa}
  M.~Herranen, T.~Markkanen and A.~Tranberg,
  %``Quantum corrections to scalar field dynamics in a slow-roll space-time,''
  JHEP {\bf 1405} (2014) 026
  [arXiv:1311.5532 [hep-ph]].
  %%CITATION = ARXIV:1311.5532;%%
	
	
	%\cite{Garbrecht:2013coa}
\bibitem{Garbrecht:2013coa} 
  B.~Garbrecht, G.~Rigopoulos and Y.~Zhu,
  %``Infrared Correlations in de Sitter Space: Field Theoretic vs. Stochastic Approach,''
  Phys.\ Rev.\ D {\bf 89}, 063506 (2014)
  [arXiv:1310.0367 [hep-th]].
  %%CITATION = ARXIV:1310.0367;%%
  %6 citations counted in INSPIRE as of 10 Dec 2014	
	
	%\cite{Serreau:2013koa}
\bibitem{Serreau:2013koa} 
  J.~Serreau,
  %``Nonperturbative infrared enhancement of non-Gaussian correlators in de Sitter space,''
  Phys.\ Lett.\ B {\bf 728}, 380 (2014)
  [arXiv:1302.6365 [hep-th]].
  %%CITATION = ARXIV:1302.6365;%%
  %8 citations counted in INSPIRE as of 10 Dec 2014

	
	%\cite{Serreau:2013eoa}
\bibitem{Serreau:2013eoa}
  J.~Serreau,
  %``Renormalization group flow and symmetry restoration in de Sitter space,''
  Phys.\ Lett.\ B {\bf 730} (2014) 271
  [arXiv:1306.3846 [hep-th]].
  %%CITATION = ARXIV:1306.3846;%%
  %4 citations counted in INSPIRE as of 10 Dec 2014

	
	%\cite{Gautier:2013aoa}
\bibitem{Gautier:2013aoa} 
  F.~Gautier and J.~Serreau,
  %``Infrared dynamics in de Sitter space from Schwinger-Dyson equations,''
  Phys.\ Lett.\ B {\bf 727}, 541 (2013)
  [arXiv:1305.5705 [hep-th]].
  %%CITATION = ARXIV:1305.5705;%%
  %10 citations counted in INSPIRE as of 10 Dec 2014
	
	%\cite{Serreau:2013psa}
\bibitem{Serreau:2013psa} 
  J.~Serreau and R.~Parentani,
  %``Nonperturbative resummation of de Sitter infrared logarithms in the large-N limit,''
  Phys.\ Rev.\ D {\bf 87}, no. 8, 085012 (2013)
  [arXiv:1302.3262 [hep-th]].
  %%CITATION = ARXIV:1302.3262;%%
  %14 citations counted in INSPIRE as of 10 Dec 2014

%\cite{Jatkar:2011ju}
\bibitem{Jatkar:2011ju} 
  D.~P.~Jatkar, L.~Leblond and A.~Rajaraman,
  %``On the Decay of Massive Fields in de Sitter,''
  Phys.\ Rev.\ D {\bf 85}, 024047 (2012)
  [arXiv:1107.3513 [hep-th]].
  %%CITATION = ARXIV:1107.3513;%%
  %20 citations counted in INSPIRE as of 10 Dec 2014
	
	%\cite{Garbrecht:2011gu}
\bibitem{Garbrecht:2011gu} 
  B.~Garbrecht and G.~Rigopoulos,
  %``Self Regulation of Infrared Correlations for Massless Scalar Fields during Inflation,''
  Phys.\ Rev.\ D {\bf 84}, 063516 (2011)
  [arXiv:1105.0418 [hep-th]].
  %%CITATION = ARXIV:1105.0418;%%
  %30 citations counted in INSPIRE as of 10 Dec 2014

\end{thebibliography}
\end{document}